\documentclass[twocolumn]{aastex631}

\usepackage{bm}% bold math
\usepackage{enumerate}% http://ctan.org/pkg/enumerate

\defcitealias{2010MNRAS.401..977J}{JT10}	

\shorttitle{Binaries and the Bar}
\shortauthors{Yavetz}

\begin{document}

\title{Disrupted Wide Binaries as Dynamical Probes of Galactic Structure}

% \correspondingauthor{Tomer Yavetz}
\email{tyavetz@ias.edu}

\author[0000-0001-6952-5364]{Tomer D. Yavetz}
\affiliation{Institute for Advanced Study, Einstein Drive, Princeton, NJ 08540, USA}

\begin{abstract}

Many of the stars in the Galaxy were formed in binary systems. The widest of these can eventually become disrupted due to a combination of kicks from passing stars and the Galactic tidal field. If the Galactic disk were purely axisymmetric, the stars from a disrupted binary system would slowly drift apart on nearly identical orbits. We study how the existence of non-axisymmetric structures, such as a rigidly rotating bar, can greatly alter this picture. In particular, we show how the orbital dynamics near the resonances sourced by these perturbations can create local fluctuations in the distribution of disrupted binary separations. We simulate the evolution of wide binary systems embedded in gravitational potentials with rotating galactic bars, and show how features and fluctuations in the distribution of disrupted binaries can be used to locate bar resonances and constrain the bar's pattern speed and amplitude.

\end{abstract}


%%%%%%%%%%%%%%%%%%%%%%%%%%%%%%%%%%%%%%%%%%%%%%%%%%

%%%%%%%%%%%%%%%%% BODY OF PAPER %%%%%%%%%%%%%%%%%%

\section{Introduction}
\label{sec:intro}

The Milky Way is generally understood to have a rigidly rotating bar-like configuration of stars at its center \citep{2016ARA&A..54..529B}. The Galactic Bar is responsible for a variety of known effects, including radial migration \citep{2010ApJ...722..112M}, separation of stellar populations in the Galactic Disk \citep{2017MNRAS.469.1587D}, and fluctuations in the density of stars in velocity space observed in the Gaia data \citep{2019A&A...626A..41M,2020MNRAS.495..895B}. The evolution of the bar may also be indicative of the properties of the Milky Way's dark matter halo \citep{2022MNRAS.513..768C}. Constraining the bar's pattern speed, amplitude, and angle with respect to the Solar neighborhood can thus shed much light on how the Milky Way evolved into its current state.

Identifying the precise characteristics of the Galactic Bar is also, unfortunately, a challenging task. The literature on this topic can be broadly divided into two types. The first aims to directly study characteristics of the bar, such as its pattern speed and its length, through either gas or stellar kinematics \citep{1975ApJ...195..617P,1991ApJ...379..631B,1991MNRAS.252..210B,2019MNRAS.488.4552S,2019MNRAS.490.4740B,2024arXiv240207986H}. The second involves studying the influence of the bar on local kinematic structures through resonant interactions \citep{2000AJ....119..800D, 2012A&A...548A.126M, 2019MNRAS.490.1026H, 2021MNRAS.500.4710C, 2021MNRAS.505.2412C}. Many of the recent studies place the pattern speed of the bar at roughly $\Omega_\mathrm{p} \approx 35-40$ km s$^{-1}$ kpc$^{-1}$, though competing claims also continue to appear periodically \citep[see for example][]{2024arXiv240207986H}. Other characteristics of the bar, including its length, mass, and pitch angle with respect to the Sun, are similarly still debated in the literature.

The dynamics of phase-mixing in the vicinity of a resonance offers up an additional, complementary method of constraining the characteristics of the bar. A rotating bar with a constant pattern speed creates a series of resonances in the local phase space around it. Primary among these are the corotation resonance and the inner and outer Lindblad resonances \citep{2008gady.book.....B}. The rate of phase-mixing in a given region of phase space depends on the local frequency gradient, and therefore, the phase-mixing process can be significantly modified near resonant regions \citep{2021MNRAS.501.1791Y,2023ApJ...954..215Y}.

The variation of phase-mixing timescales can thus help in identifying resonant regions, and by extension, probing galactic structure. Leveraging this requires that the initial distribution of tracer particles not be initially phase-mixed, and conveniently, the initial conditions of the Galactic Disk (and the halo) are anything but phase-mixed. Star formation is a localized process, and stars are preferentially born together, whether in large globular clusters, medium-sized open clusters in the Galactic Disk, and/or as part of binary systems. As these star associations evolve in the Galactic tidal field, they are gradually unbound due to a combination of dynamical processes. Once they are unbound, their subsequent evolution is well-characterized as a collection of massless test particles on similar orbits \citep{2019ApJ...884L..42K}.

Binary systems in particular are ubiquitous throughout the Galactic Disk, accounting for over 50\% of the Galaxy's stars by some estimates \citep{2024NewAR..9801694E}. Of specific interest are very weakly bound binary systems with separations $\gtrsim0.1$ pc \citep[see, e.g.,][]{2017arXiv170903532P,2018MNRAS.476..528E,2021MNRAS.501.3670P,2022MNRAS.512.3383H}, because these are most sensitive to the Galactic potential. \citet{2010MNRAS.401..977J} \citepalias[henceforth][]{2010MNRAS.401..977J} studied the evolution of wide binary systems in the Galactic tidal field, with an emphasis on understanding how the Galactic tidal field affects the distribution of stars both before and after the disruption of the binary system. By tracking the evolution of the unbound stars, \citetalias{2010MNRAS.401..977J} found that they could reach separations of up to several hundred parsecs over the age of the Galaxy. They argued that this unbound subset can introduce correlations in the positions and velocities of disk stars on small-intermediate scales that could be observed in surveys such as Gaia.

\citetalias{2010MNRAS.401..977J} focused their investigation on binary systems in the Solar neighborhood, assuming an axisymmetric tidal field for the Galaxy. As mentioned above, the existence of non-axisymmetric perturbations such as the Galactic Bar introduces resonances that can greatly alter the evolution of this subset of stars from disrupted binary systems. The central focus of this work is thus to extend \citetalias{2010MNRAS.401..977J} in order to study how the distribution of unbound binary systems varies across the Galactic Disk, and how it responds to the dynamical influence of a resonance. The ultimate aim is to enable the identification of unique variations across the Galactic Disk that can be used to pinpoint the location of resonances due to Galactic structures such as the bar. 

This paper is structured as follows: Section \ref{sec:theory} lays out the theoretical basics for analyzing the orbital dynamics near resonances and introduces a toy model for understanding how phase-mixing is altered near a resonance. Section \ref{sec:binaries} reviews the ingredients necessary for studying the evolution and disruption of binary systems in the presence of the Galactic tidal field as well as kicks from passing stars (largely following the treatment in \citetalias{2010MNRAS.401..977J}). In Section \ref{sec:sims}, we simulate the evolution of binary systems embedded in non-axisymmetric galactic potentials, and study the variations in the distribution of binary separations in different regions of the disk. Finally, we discuss shortcomings of this approach and lay out avenues for detecting the variation signal in Section \ref{sec:disc}.

\section{Theory}
\label{sec:theory}

In this section, we review a few standard concepts that are central to the study of near-resonant dynamics, and use them to lay the groundwork for understanding why disrupted binary systems are helpful for the identification of resonant regions in the Milky Way's disk. In Section \ref{subsec:resonances}, we discuss the use of the pendulum Hamiltonian as an analogy for the motion of stars near a resonance with the Galactic Bar. In Section \ref{subsec:toymodel}, we build and analyze a toy model that demonstrates the phase-mixing process that governs the results described in the subsequent sections.

\subsection{Orbital Dynamics near a Resonance}
\label{subsec:resonances}

The aim of this section is to characterize the motion of a test particle in a potential with a rotating non-axisymmetric perturbation, such as a rigidly rotating quadrupole-like bar embedded in an axisymmetric disk+halo potential. As in the unperturbed axisymmetric case, most orbits are still well-described through the use of `angle-action' coordinates, where the actions $\bm{J} = (J_R, J_\phi, J_z)$ represent adiabatic invariants of the motion, and the conjugate angles $\bm{\theta} = (\theta_R, \theta_\phi, \theta_z)$ represent the phase of the orbit. The Hamiltonian of the particle is:
\begin{equation}
    \label{eq:ham}
    H(\bm{\theta},\bm{J}) = H_0(\bm{J}) + \delta \Phi(\bm{\theta},\bm{J}) \ ,
\end{equation}
where $H_0$ represents the unperturbed, mean-field Hamiltonian, and $\delta\Phi$ is a small perturbation.

Since the actions are adiabatic invariants, Hamilton's equations dictate that the angles evolve linearly with time at constant frequencies:
\begin{equation}
    \label{eq:ham_freq}
    \bm{\Omega}(\bm{J}) \equiv \frac{\partial H_0}{\partial \bm{J}} = (\Omega_R, \Omega_\phi, \Omega_z) \ .
\end{equation}

The non-axisymmetric perturbation causes the actions to fluctuate, though in most locations these fluctuations are small and can be neglected for all practical purposes. The fluctuations become important in the vicinity of orbital resonances, defined as locations where
\begin{equation}
    \label{eq:resonance}
    \bm{N}\cdot\bm{\Omega} = N_\phi\Omega_\mathrm{p} \ ,
\end{equation}
for a rigidly rotating perturbation with a pattern speed $\Omega_\mathrm{p}$. $\bm{N}$ represents some vector of integers $(N_R, N_\phi, N_z)$ that satisfies the resonance condition. A straightforward example of this is the \textit{corotation} resonance, where $\bm{N} = (0,m,0)$.

Analyzing near-resonant orbits requires the use of secular perturbation theory \citep{1992rcd..book.....L}. By converting to a new set of canonical coordinates defined by the resonant condition, one can separate the problem into `fast' and `slow' coordinates, with the latter characterizing the slow motion with respect to the stable equilibrium of the resonance:
\begin{equation}
    \label{eq:slow_action-angle}
    \theta_\mathrm{s} \equiv \bm{N}\cdot\bm{\theta} - N_\phi\Omega_\mathrm{p} t \quad \mathrm{and} \quad J_\mathrm{s} \equiv \frac{J_\phi}{N_\phi} \ .
\end{equation}

The fast angles evolve on the orbital timescale, whereas the slow angle evolves considerably more slowly in the vicinity of the resonance. As a result, averaging the Hamiltonian over the fast angles allows one to reduce the problem to one degree of freedom in the slow plane. The resultant Hamiltonian takes the form of the pendulum equation:

\begin{equation}
    \label{eq:pendulum}
    h(\varphi,I) = \frac{1}{2}GI^2 - F\cos(k\varphi) \ ,
\end{equation}
with $\varphi = \theta_\mathrm{s} + C$, where $C$ is a constant related to the initial angle of the perturbing potential, and $I = J_\mathrm{s} - J_\mathrm{s,res}$, where $J_\mathrm{s,res}$ is the value of $J_\mathrm{s}$ for which Equation \ref{eq:resonance} is satisfied in the unperturbed potential. $k$ is the Fourier component that dominates perturbation $\delta\Phi$ ($k$ is typically a small integer). The constants $G$ and $F$ depend on the curvature of the unperturbed Hamiltonian and on the strength of the perturbation, respectively. We refer the reader to chapter IV, section F of \citet{2024arXiv240213322H} for the full derivation of Equation \ref{eq:pendulum}, as well as the precise definitions of $G$ and $F$.

\begin{figure}
    \includegraphics[width=\columnwidth]{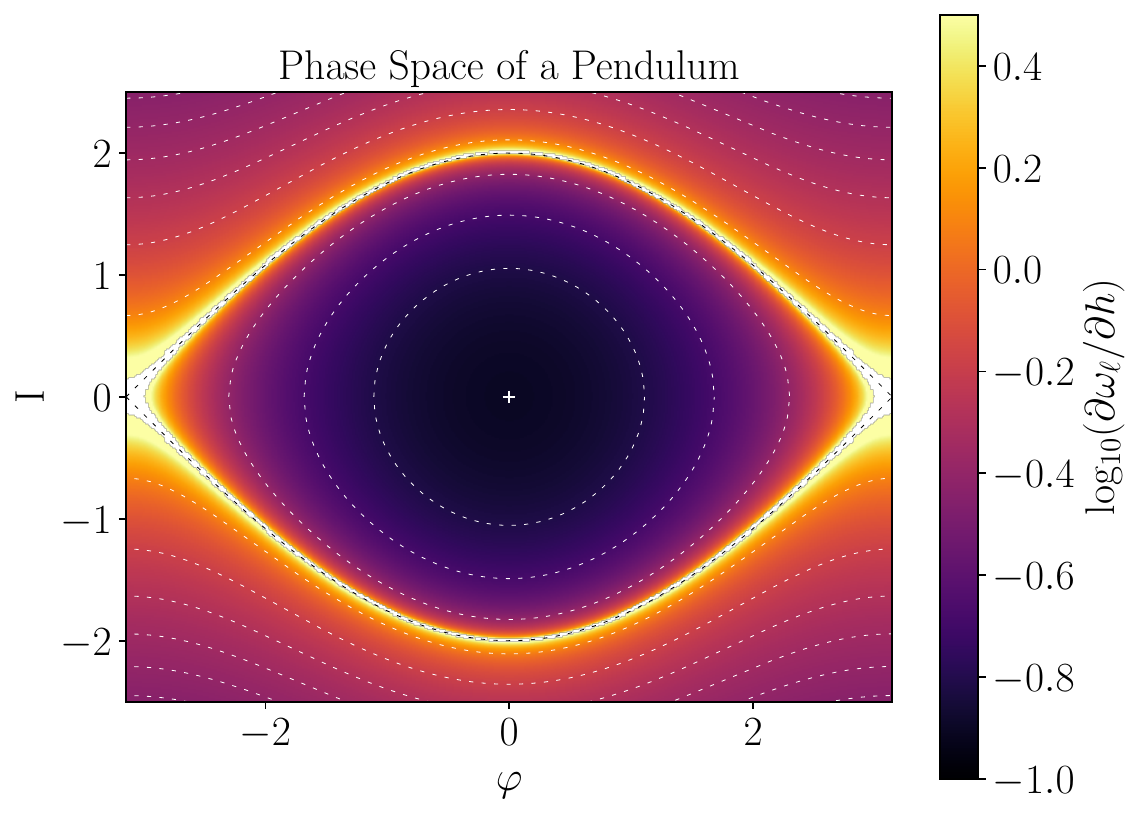}
    \caption{The phase space of the pendulum defined in Equation \ref{eq:pendulum}, with $k=G=F=1$. The dashed white lines represent contours of equal $h$, the stable equilibrium (where $h=-F$) is marked with a white $+$, and the separatrix (where $h=F$) is represented by the dashed black lines. The color at each location in the phase space diagram represents the log value of the gradient of the pendulum's frequency with respect to the Hamiltonian, revealing a plateau at the stable equilibrium and a sharp discontinuity at the separatrix.}
    \label{fig:01_Phase_Space_of_a_Pendulum}
\end{figure}

The phase space of the pendulum, defined by Equation \ref{eq:pendulum} with $k=1$ and shown in Figure \ref{fig:01_Phase_Space_of_a_Pendulum}, provides a helpful framework for studying orbits near the corotation resonance with a rotating galactic structure such as a rigid bar. Each contour representing an orbit defined by the value of the Hamiltonian $h$ along that given orbit. Several noteworthy characteristics of the resonant dynamics are readily apparent in this figure: a stable equilibrium is found at $(\varphi,I) = (0,0)$, where $h=-F$. Surrounding this point is a family of orbits that are unable to access the full range of $\varphi$; instead, they are \textit{trapped} around the stable equilibrium. These are referred to as \textit{librating} orbits. In the case of the Galactic Bar, these are stellar orbits that remain on one side of the bar (in the corotating frame with the bar). Only when $h\geq F$ do orbits explore the full range of $\varphi$ -- such orbits are typically referred to as \textit{circulating} orbits. The boundary between these two orbit families, where $h=F$, is known as the \textit{separatrix}, and the special point at $(\varphi,I) = (\pi, 0)$, where the lines of the separatrix cross, is the unstable equilibrium.

The size of the trapped region is an indicator of the strength of the perturbation, and is broadly defined using the half-width of the orbit family:
\begin{equation}
    \label{eq:half-width}
    I_\mathrm{h} \equiv 2\sqrt{\frac{F}{G}} \ .
\end{equation}

The generalized libration frequency of a trapped orbit within this part of the pendulum phase space is:
\begin{equation}
    \label{eq:libration_freq}
    \omega_\ell(h) = \pi\sqrt{2GF}\bigg[\int_{-\varphi_0}^{\varphi_0} \frac{d\varphi}{\sqrt{\cos\varphi - \cos\varphi_0}}\bigg]^{-1} \ ,
\end{equation}
where $\varphi_0$ represents the maximum value of $\varphi$ (i.e., where $I = 0$) for an orbit defined by its value of $h$:
\begin{equation}
    \label{eq:varphi_0}
    \varphi_0(h) = \arccos(-h/F) \ .
\end{equation}

Orbits with small $\varphi_0$ remain close to the stable equilibrium, and in these cases Equation \ref{eq:pendulum} approximates to the Hamiltonian of a harmonic oscillator. The libration frequency in this region is roughly constant at:

\begin{equation}
    \label{eq:res_freq}
    \omega_{\ell,\mathrm{res}} \sim \sqrt{GF} \ .
\end{equation}

As one approaches the separatrix, $\omega_\ell$ goes to zero (the libration period diverges to infinity). On the other side of the separatrix, the circulation frequency is given by:
\begin{equation}
    \label{eq:circulation_freq}
    \omega_\ell(h) = \pi\sqrt{8GF} \bigg[\int_{-\pi}^\pi \frac{d\varphi}{h/F + \cos\varphi}\bigg]^{-1} \ .
\end{equation}

Figure \ref{fig:01_Phase_Space_of_a_Pendulum} is colored based on the gradient of $\omega_\ell$ at each location on the plot. Unsurprisingly, the coloring follows the contours of constant $h$, given that $\omega_\ell$ as defined above is only a function of $h$. However, we note that while $h$ grows monotonically from $(0,0)$ outward, $\omega_\ell$ is initially constant in the vicinity of $(0,0)$, and subsequently exhibits a discontinuity at the separatrix.

In theory, identifying the dynamical characteristics of a range of orbits in a given system would make pinpointing the location and width of the resonance a simple exercise. In practice, however, we are only afforded a single snapshot of a local patch of our Galaxy, without a clear path to identifying invariant dynamical quantities such as the actions or frequencies of a given star. In the next subsection, we demonstrate how the existence of test particles with similar initial conditions can be utilized to overcome this challenge.

\begin{figure}
    \includegraphics[width=\columnwidth]{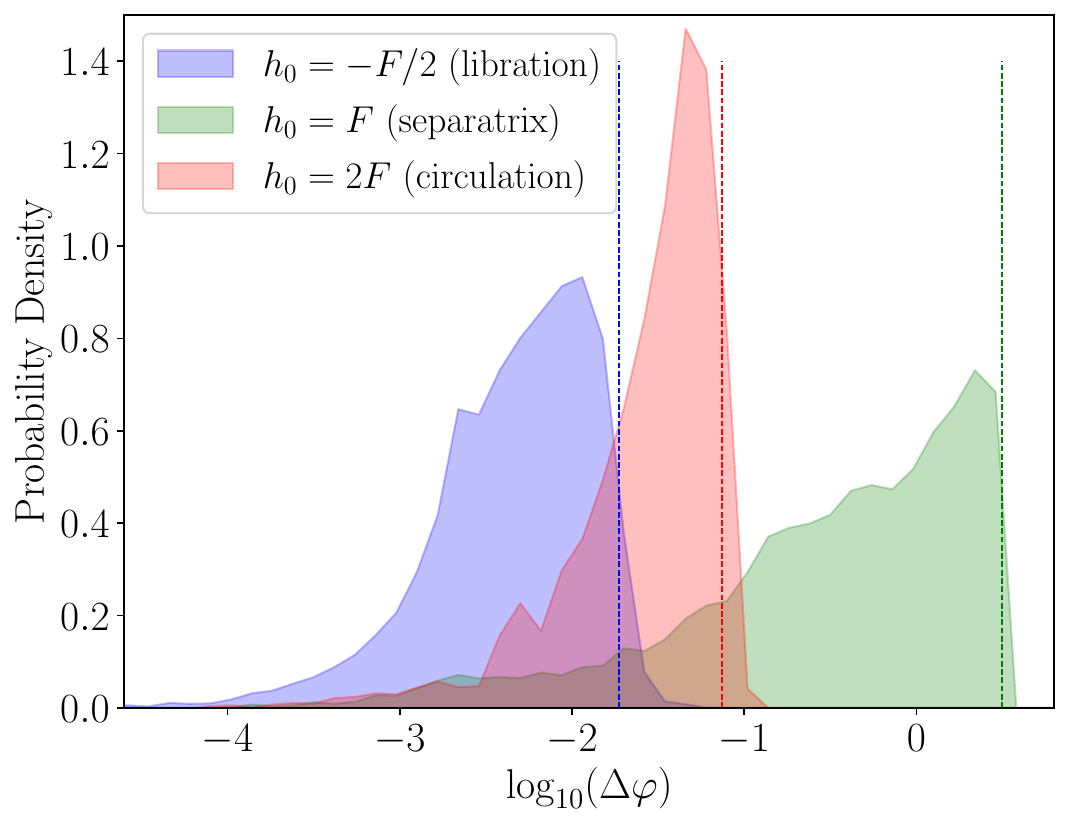}
    \caption{Histograms of the separations of three ensembles of $10^5$ pairs of pendula, initialized at three values of $h_0$, corresponding to different types of pendulum motion, and evolved for 10 resonant libration times (see Equation \ref{eq:res_freq}). Pendula in the librating regime (in blue) remain closer to each other throughout than pendula in the circulating regime (in red). Twin pendula initialized near the separatrix (in green) attain much greater separations, up to the maximum size of the system ($\varphi=\pi$). The dashed lines correspond to the predicted maximum separation for each regime based on Equation \ref{eq:varphi_max_est}.}
    \label{fig:02_delta_varphi_prob}
\end{figure}

\subsection{Twin Pendulum Toy Model for Binaries Near a Resonance}
\label{subsec:toymodel}

Consider the following scenario: an ensemble of pendula is initialized with a range of values of $h$ in the $(\varphi,I)$ plane (equivalent to a collection of stars orbiting in the Galactic Disk in the vicinity of a resonance with the Galactic Bar). A single snapshot of the $(\varphi,I)$ values at an arbitrary time sheds little light on the characteristics of the system. One can overcome this by having access to a second snapshot from a different time -- e.g., by tagging each pendulum with its initial conditions\footnote{This is the approach underlying several recent studies of the Galactic Bar, that utilize the chemical abundances or the ages of observed stars to estimate their initial conditions \citep[see, e.g.,][]{2021MNRAS.505.2412C,2022ApJ...935...28W}.}. However, we assume for the time being we have no way of tracing each pendulum (or star) back to its initial conditions.

Now, suppose that each pendulum with energy $h_0$ is known to have started out with a nearby companion pendulum with the same initial value of $\varphi$ and $h = h_0+\delta h$. Following equations \ref{eq:libration_freq} and \ref{eq:circulation_freq} and the structure shown in Figure \ref{fig:01_Phase_Space_of_a_Pendulum}, the frequency difference between the two pendula will greatly depend on the value of the energy $h_0$ (and on the characteristics of the system). As a consequence, the rate at which the pendula drift apart from each other will also greatly depend on $h_0$. The separation between two pendula with $h_0$ and $h_0+\delta h$ after an arbitrary period of time $t$ can be estimated based on the dependence of the frequency on the Hamiltonian:
\begin{equation}
    \label{eq:varphi_max_est}
    \Delta\varphi(t) \approx \frac{\partial\omega_\ell}{\partial h}\bigg|_{h_0} \times \delta h \times t \ .
\end{equation}

This setup is not altogether contrived; most stars in the Galaxy are born in binary systems \citep{2024NewAR..9801694E}, and though these systems are initially gravitationally bound, many of them are disrupted over time, at which point the evolution of the two stars is qualitatively similar to that of two test particles initialized at similar phases on nearby orbits.

In practice, $\Delta\varphi$ in Equation \ref{eq:varphi_max_est} does not grow monotonically, and depends on the precise location of the stars in the $(\varphi,I)$ plane ($\Delta\varphi$ reaches a local maximum when $\varphi=0$, and a local minimum when $I=0$ for librating orbits or $\varphi=\pi$ for circulating orbits). The eventual separations of twin pendula (or two stars from a disrupted binary system) separated by $\delta h$ is therefore highly sensitive to the location of the resonance and its characteristics. Figure \ref{fig:02_delta_varphi_prob} depicts histograms of the final separations $\Delta\varphi$ of ensembles of $10^5$ twin pendula initialized at three different energies: $h_0=\{-F/2, F, 2F\}$. One pendulum from each pair is randomly initialized in the $(\varphi,I)$ plane, and its twin is initialized at the same $\varphi$, with $I$ set such that $h = h_0 \pm \delta h$ with $\delta h = 0.002F$. The two pendula are then integrated for ten resonant libration periods ($t_\mathrm{int} = 10\times2\pi/\omega_{\ell,\mathrm{res}} = 20\pi/\sqrt{FG}$), at which point the separation in $\varphi$ between each pair of pendula is recorded. The pendula initialized at $h=-F/2$ (plotted in blue) are well within the harmonic oscillator regime, meaning each pendulum pair has nearly identical frequencies, and $\Delta\varphi$ remains small throughout their evolution. The opposite is true for the pendula initialized at the separatrix (plotted in green), whose separations grow large enough to attain the maximum possible separation of $\Delta \varphi = \pi$ within the integration time. The final separations of circulating pendula (plotted in red), which possess intermediate values of $\partial\omega_\ell / \partial h |_{h_i}$, fall in between these two extremes, as one might expect from Equation \ref{eq:varphi_max_est} and Figure \ref{fig:01_Phase_Space_of_a_Pendulum}. The dashed blue and red lines in Figure \ref{fig:02_delta_varphi_prob} correspond to the respective predictions from Equation \ref{eq:varphi_max_est}, and accurately predict the maximal separation of each ensemble. The dashed green line is at $\Delta\varphi=\pi$ (per Equation \ref{eq:varphi_max_est} the value of $\Delta\varphi$ at the separatrix is infinite, and the maximal separation in this setup is therefore limited to $\pi$).

\begin{figure}
    \includegraphics[width=\columnwidth]{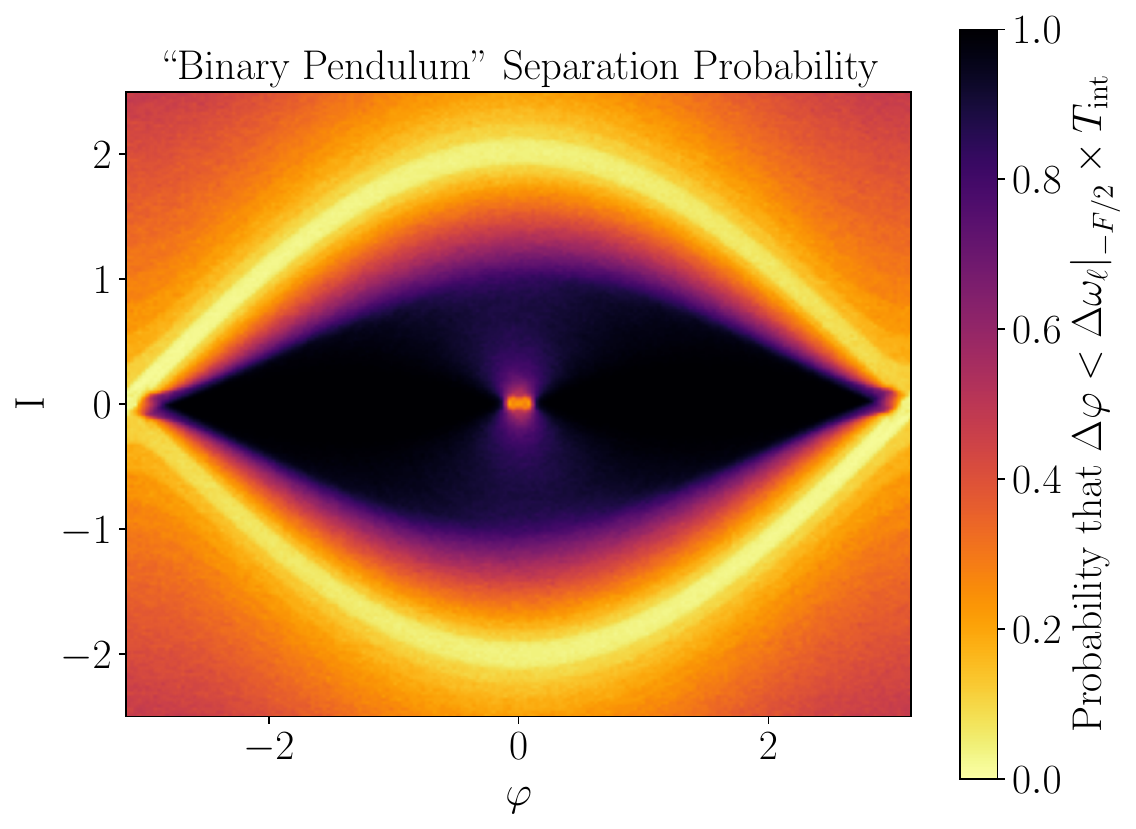}
    \caption{The phase space of the same pendulum shown in Figure \ref{fig:01_Phase_Space_of_a_Pendulum}, except here each pixel is colored based on the probability that a pendulum in that location in $(\varphi, I)$ remains separated by less than a certain value of $\Delta\varphi$ from its twin. The value of $\Delta\varphi$ chosen here is determined using Equation \ref{eq:varphi_max_est} evaluated at $h_0=-F/2$, also corresponding to the blue dashed line in Figure \ref{fig:02_delta_varphi_prob}.}
    \label{fig:03_Phase_Space_of_a_Pendulum_from_binary_separations}
\end{figure}

The dependence of twin pendulum separation at each location in the $(\varphi, I)$ plane on the properties of the system can thus enable one to deduce those properties from a single snapshot of phase space, assuming some knowledge about which pendula started out as twins (but without needing to know any of their initial conditions besides the fact that they started out close to each other in phase space). This is shown in Figure \ref{fig:03_Phase_Space_of_a_Pendulum_from_binary_separations}, where the setup and the contours are identical to those of Figure \ref{fig:01_Phase_Space_of_a_Pendulum}, but the color now represents the probability that $\Delta\varphi$ between the twin pendula at each location remains below a certain value, chosen here to be the value determined by Equation \ref{eq:varphi_max_est} evaluated at $h_0=-F/2$. In other words, the color represents the area under the histograms in Figure \ref{fig:02_delta_varphi_prob} up to the blue dashed line. Each set of pendula is initialized as described above, and evolved for ten times the resonant libration period, as before. The same structure from Figure \ref{fig:01_Phase_Space_of_a_Pendulum} is immediately apparent in this figure, with pendula in the librating zone remaining close to one another, and pendula at the separatrix attaining the largest separations. 

Figures \ref{fig:01_Phase_Space_of_a_Pendulum} and \ref{fig:03_Phase_Space_of_a_Pendulum_from_binary_separations} are not identical to one another, and the existence of twin pendula at small separations at almost all points along $I=0$ stands out in contrast to the contours of constant $h$ in Figure \ref{fig:01_Phase_Space_of_a_Pendulum}. This feature is present due to the nature of the motion defined by Equation \ref{eq:pendulum}, whereby the librating pendula slow down as $|\varphi|$ reaches its maximum, allowing a lagging pendulum to temporarily `catch up' to the leading pendulum. The small bright spot at $(\varphi, I) = (0,0)$ is a consequence of the choice of initial conditions when $\delta h$ is comparable to $h_0$, and is irrelevant to the results discussed below.

The introduction of a resonance due to a perturbation in the Galaxy thus creates two unique regimes, that do not exist in the unperturbed case, in terms of the final separations between two stars that start out close to each other. The separation between stars on trapped orbits in the harmonic region grows very slowly, whereas in the vicinity of the separatrix, the twin stars diverge from each other rapidly, especially when viewed far enough away from the unstable equilibrium. Both of these regimes stand in contrast to regions unaffected by the resonance, where the eventual separation in angle space can be predicted simply through $\Delta\bm{\theta} = \Delta\bm{\Omega}t$ in the unperturbed potential.

In the next section, we describe our method for modeling the disruption of binary systems and their subsequent evolution in the Galactic potential, and in Section \ref{sec:sims} we put these two pieces together to model how the existence of a rotating bar changes the distribution of disrupted binary star separations at different locations in the Galactic Disk.

\section{Evolution of Wide Binaries}
\label{sec:binaries}

Understanding the evolution of wide binary systems embedded in the Galactic tidal field is central to the aims of this work. In particular, the ability to identify the presence of a resonance depends on the predicted distribution of disrupted wide binary separations in the unperturbed case, as postulated in the previous section. In this section, we lay out our method for studying the evolution of binary systems before and after their disruption, which follows the treatment in \citetalias{2010MNRAS.401..977J} to a large extent, with a few notable exceptions\footnote{A similar setup is described in \citet{2024arXiv240502912S}, though the focus there is on how galactic tides drive wide binary systems to close encounters and mergers, rather than on the evolution of disrupted pairs.}. In general, the key ingredients of our method are (1) assumptions regarding the initial distribution of wide binaries, (2) a prescription for how they evolve as a restricted three-body system (i.e., two gravitating stars in the presence of the Galactic gravitational field), and (3) an algorithm for modeling the effect of perturbations from passing stars, that can eventually cause some of the binary systems to become unbound. We describe each of these ingredients briefly in what follows, and refer the reader to \citetalias{2010MNRAS.401..977J} for a more thorough description.

\subsection{Binary System Initial Conditions}
\label{subsec:binary_inits}

The initial conditions of a binary system are fully specified through 6 orbital parameters: the semi-major axis ($a$), the eccentricity ($e$), the inclination of the orbit ($i$), the longitude of the ascending node ($\Omega_\mathrm{o}$), the argument of periapsis ($\omega_\mathrm{o}$), and the mean anomaly\footnote{We append the subscript `o' to some of the orbital elements to avoid confusion with the notation of frequencies from the previous section}. We assume $\Omega_\mathrm{o}$, $\omega_\mathrm{o}$, and the mean anomaly are uniformly distributed between $0$ and $2\pi$, and $\cos i$ is uniformly distributed between $-1$ and $1$. We define the inclination of the orbit with respect to the plane of the Galactic Disk, and the longitude of the ascending node with respect to the direction to the Galactic center.

We assume an initially thermal distribution\footnote{The eccentricity distribution of wide binaries in the Milky Way is actually understood to be somewhat superthermal \citep{2022MNRAS.512.3383H}. While we initialize all of our simulations below by drawing from a thermal distribution, we find that changing the eccentricity distribution to mildy sub- or superthermal has no effect on our results.} of eccentricities, meaning $e^2$ is uniformly distributed between $0$ and $1$. Finally, we follow \"{O}pik's Law for the distribution of semi-major axes (i.e., a log-uniform distribution), ranging between $0.002$ pc and $1$ pc.

As a simplifying assumption, we take all of our systems to be equal-mass binaries with solar mass stars, such that the reduced mass of each system is $\mu=0.5 \mathrm{M}_\odot$ and the Jacobi radius $r_J = 1.7$ pc. We further assume that the birth times of these binaries are evenly distributed between the present day and 10 Gyr ago, equivalent to the \textit{\"{O}pik 1} setup described in \citetalias{2010MNRAS.401..977J}.

\subsection{Binary System Evolution in the Galactic Potential}
\label{subsec:binary_evol}

In order to track the evolution of a binary system embedded in the Galactic Disk, \citetalias{2010MNRAS.401..977J} make use of Hill's approximation and derive the equations of motion for binary systems in the Solar neighborhood, assuming the systems follow roughly circular galactic orbits in an axisymmetric galaxy. Given our primary goal is to study how binary systems evolve in different regions of non-axisymmetric galactic potentials, we instead choose to numerically integrate the orbits of the two stars under the influence of their own gravitational fields superposed with an arbitrary Galactic gravitational potential, with the option of including non-axisymmetric features in this external potential such as the perturbation sourced by a rotating bar.

We utilize the \texttt{gala} code suite to run the simulations described in Section \ref{sec:sims}, and in particular we rely on the \texttt{gala.dynamics.nbody} package to model a binary system embedded in the Galactic Disk. The integration is carried out with the Dormand-Prince 8(5,3) integration scheme (an eighth order Runge-Kutta method) using an adaptive timestep to ensure that both the binary orbit and the Galactic orbit are well resolved at every step. In practice, we choose the timestep to be the lesser of $0.1$ Myr and $0.1 / \Omega_\mathrm{b}$, with:
\begin{equation}
    \label{eq:binary_frequency}
    \Omega_\mathrm{b} \equiv \sqrt{\frac{G(M_1 + M_2)}{a^3}} \ .
\end{equation}
We verify that energy is conserved to within numerical precision between the application of kicks (see next subsection), and further reductions of the integration timestep do not alter our results.

For the background axisymmetric potential, we utilize the \texttt{MilkyWayPotential2022} from \texttt{gala}, which provides an excellent fit to recent data, particularly in the inner 15 kpc of the Milky Way, including the \citet{2019ApJ...871..120E} rotation curve (all of our simulated binary systems remain within the inner $15$ kpc for the duration of the simulation). This potential model consists of a nucleus and a bulge, both modeled using spherical Hernquist potentials, a spherical NFW halo, and a 3-component sum of Miyamoto-Nagai disks constructed to represent an exponential disk. More detail, including the specific parameters for each of the potential's components, can be found in the \texttt{gala} online documentation\footnote{\url{https://gala.adrian.pw/en/latest/}}.

In order to study the effects of the Galactic Bar, we add a quadrupolar ($m=2$) perturbation of the following form:
\begin{equation}
    \label{eq:quad_bar}
    \delta\Phi(R,\phi,z) = \Phi_\mathrm{b}(R)\frac{R^2}{R^2+z^2}\cos\big[m(\phi-\Omega_\mathrm{p}t)\big] \ ,
\end{equation}

\begin{equation}
    \label{eq:bar_amp}
    \Phi_\mathrm{b}(R) = -\frac{Av_\mathrm{c}^2}{m}\bigg[\frac{R}{R_\mathrm{CR}}\bigg]^2\bigg[\frac{b+1}{b+R/R_\mathrm{CR}}\bigg]^5 \ ,
\end{equation}
with $\Omega_\mathrm{p} = 40$ km s$^{-1}$ kpc$^{-1}$, $A = 0.02$, $v_c = 230$ km s$^{-1}$, and $R_\mathrm{CR} = 5.75$ kpc.

\begin{figure}
    \includegraphics[width=\columnwidth]{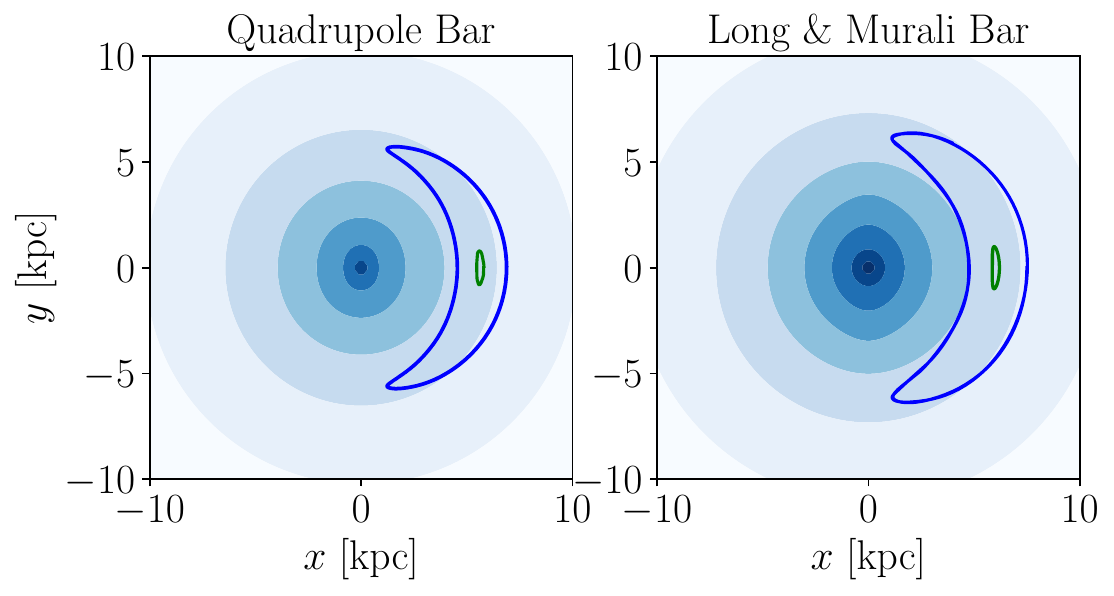}
    \caption{Isopotential contours of the two Galactic potentials used in Section \ref{sec:sims}. In each case, two trapped orbits are also shown (plotted in the co-rotating frame with the bars, $\Omega_\mathrm{p} = 40$ km/s/kpc); one that is close to the stable equilibrium (in green) and another that is close to the separatrix (in blue).}
    \label{fig:04_pot_comp}
\end{figure}

To further generalize our results, we also consider a second prescription for the potential of the Galactic Bar based on \citet{1992ApJ...397...44L} with $M=10^{10}$ M$_\odot$, $a=4$ kpc, $b=0.8$ kpc, and $c=0.25$ kpc. Figure \ref{fig:04_pot_comp} shows of comparison of the density contours of the two barred potentials described above, with a few special orbits plotted in each case. The Long \& Murali bar involves a superposed triaxial density profile on top of the axisymmetric base potential, so the total mass of the galaxy is increased by $M=10^{10}$ M$_\odot$, and the location of co-rotation is therefore slightly further away from the galactic center compared to the quadrupole bar.

\subsection{Kicks from Passing Stars}
\label{subsec:binary_kicks}

The primary driver of the growth of a wide binary's semi-major axis (and in some cases, the eventual dissolution of the system), is the existence of other perturbers that kick the stars of the binary system away from their original orbit. These kicks are typically small and only cause minor changes to the various orbital parameters of the binary system, but the cumulative effect of many such encounters leads to considerable changes to the binary's orbit over time \citep{1987ApJ...312..367W}.

As in \citetalias{2010MNRAS.401..977J}, we consider only stellar encounters, and ignore any additional effects from other perturbers such as molecular clouds or dark matter substructure. Our results should therefore not be taken as the true expected distribution of disrupted wide binaries (there are various other reasons our results are only qualitatively correct, as discussed in greater detail in Section \ref{sec:disc}). The main aim of this work is indeed not to accurately calculate the true distribution of disrupted wide binaries, but rather to characterize how the distribution of disrupted wide binaries changes as a function of position with respect to a resonance.

We assume all kicks are in the impulsive regime, and therefore impart an instantaneous change in velocity to each member of the binary system without changing their positions (for sufficiently wide binaries, the velocity dispersion of stars in the disk that could perturb the binary is typically more than an order of magnitude greater than the relative velocity between the two members of the binary system). Furthermore, we rely on the central limit theorem, whereby the effect of many individual kicks that cause a cumulative change 
of $\Delta\bm{v}$ to the velocity of a subject star over a time interval $\Delta t_\mathrm{p}$, is identical to that of normal distribution with the same mean $\bm{\mu} = \langle\Delta\bm{v}\rangle$ and covariance matrix $C_{\alpha\beta} = \langle\Delta v_\alpha \Delta v_\beta\rangle$, with $\alpha$ and $\beta$ representing the different components of $\Delta\bm{v}$. In the limit where the velocity of a subject star relative to the center of mass of the binary $v = |\bm{v}| \ll \sigma$, the mean and covariance can be obtained using the diffusion coefficients, as described in \citetalias{2010MNRAS.401..977J}. In this specific setup, these reduce to:

\begin{eqnarray}
    \label{eq:diff_coeffs1}
    \mu_\alpha & = & D[v_\alpha]\Delta t_\mathrm{p} = \frac{v_\alpha}{v}D[\Delta v_\parallel]\Delta t_\mathrm{p} \ , \\
    C_{\alpha\beta} & = & D[\Delta v_\alpha \Delta v_\beta]\Delta t_\mathrm{p} = \frac{1}{2}\delta_{\alpha\beta}D[(\Delta v_\perp)^2]\Delta t_\mathrm{p} \ ,
\end{eqnarray}
with
\begin{eqnarray}
    \label{eq:diff_coeffs2}
    D[\Delta v_\parallel] & = & \frac{4\sqrt{2\pi}G^2(M\rho_1 + \rho_2) \ln{\wedge}}{3\sigma^3}v \ , \\
    D[(\Delta v_\perp)^2] & = & \frac{16\sqrt{2\pi}G^2\rho_2 \ln{\wedge}}{3\sigma} \ .
\end{eqnarray}

\begin{figure*}
    \includegraphics[width=\textwidth]{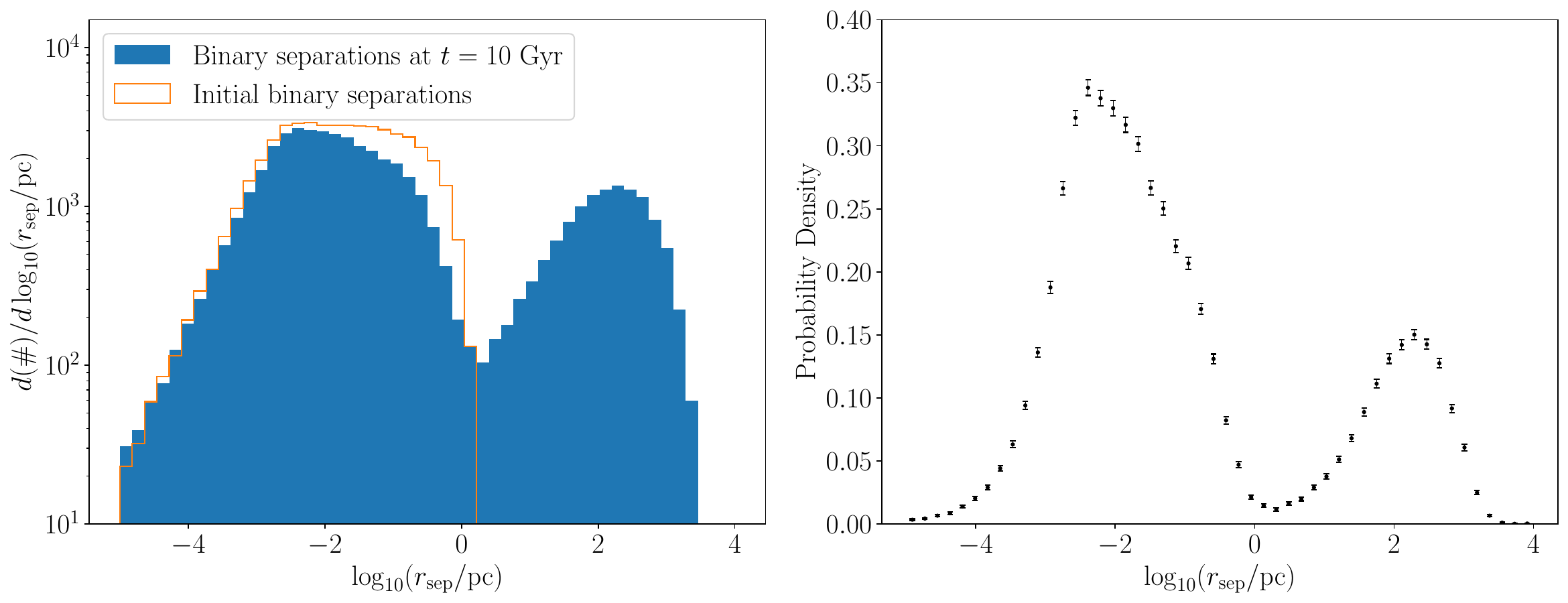}
    \caption{\textit{Left}: Reproduction of the \"{O}pik 1 simulation result in \citetalias{2010MNRAS.401..977J}, showing the initial (hollow orange histogram) and final (filled blue histogram) projected separations of $5\times 10^4$ binary systems evolved in an axisymmetric disk, on initially circular orbits at the solar neighborhood (see panel 5 of Figure 3 in \citetalias{2010MNRAS.401..977J} for comparison). The projected separations are measured from a random viewing angle for each binary system. \textit{Right}: The probability distribution function corresponding to the histogram on the left.}
    \label{fig:05_JT_reprod}
\end{figure*}

In the above, $M$ is the mass of the subject star ($M = M_1 = M_2 = \mathrm{M}_\odot$), and $\rho_1$ and $\rho_2$ represent the moments of the mass function of stars in the neighborhood of the binary system, for which we adopt the values from \citetalias{2010MNRAS.401..977J}: $\rho_1 = 0.045\ \mathrm{M}_\odot\ \mathrm{pc}^{-3}$ and $\rho_2 = 0.029\ \mathrm{M}^2_\odot\ \mathrm{pc}^{-3}$. We set:
\begin{equation}
    \wedge = \frac{b_\mathrm{max}v_\mathrm{typ}^2}{G(M+M_\mathrm{p})} \ ,
\end{equation}
where we take $b_\mathrm{max}$ (the maximum impact parameter) to be half the separation between the two stars when the kick is applied, and $v_\mathrm{typ}$ to be the local velocity dispersion $\sigma$.
As further simplifications, we set the perturber mass to be $M_\mathrm{p} = \mathrm{M}_\odot$, and $\sigma=40$ km/s, independent of location (introducing a gradient in the velocity dispersion as a function of the position in the disk has no effect on the relative behavior of wide binaries in barred vs. unbarred disk profiles).

To further accelerate our calculations, we switch to the diffusion approximation for wide binaries with sufficiently small semi-major axes. We note that our key results are primarily driven by those binary systems that start out with larger separations, for which we do not use the diffusion approximation. We refer the reader to \citetalias{2010MNRAS.401..977J}, Section 2.3 for the precise details of our implementation of the diffusion approximation.

We choose the time interval $\Delta t_\mathrm{p}$ between applying subsequent kicks based on the binary configuration. For binaries with relatively small semi-major axes, we set it to the expected time between encounters with an impact parameter smaller than the semi-major axis:
\begin{equation}
    \label{eq:Dt1}
    \Delta t_\mathrm{p} = 1.25 \times 10^7 \bigg(\frac{0.1\ \mathrm{pc}}{a}\bigg)^2 \ \mathrm{yr} \ .
\end{equation}

For larger semi-major axes, we set the time interval to be a small fraction of the orbital timescale or the Galactic timescale, whichever is smaller:

\begin{equation}
    \label{eq:Dt2}
    \Delta t_\mathrm{p} = \frac{0.1}{\mathrm{max} (\Omega_\phi,\Omega_\mathrm{b})} \ ,
\end{equation}
with $\Omega_\phi$ and $\Omega_\mathrm{b}$ defined in equations \ref{eq:ham_freq} and \ref{eq:binary_frequency}, respectively. At intermediate separations, we choose the larger of equations \ref{eq:Dt1} and \ref{eq:Dt2}. Finally, if the binary system is unbound, we continue to apply random kicks at $\Delta t_\mathrm{p} = 0.1 / \Omega_\phi$.

As a test of the setup described in Sections \ref{subsec:binary_inits}-\ref{subsec:binary_kicks}, we simulate $5\times10^4$ binaries in the axisymmetric background potential without any non-axisymmetric perturbations superimposed. All the binaries are initialized with $J_\phi = 1850$ km s$^{-1}$ kpc$^{-1}$, which is approximately equivalent to a star on a circular orbit in the Solar neighborhood assuming the \citet{2019ApJ...871..120E} rotation curve. The resulting distribution of projected separations, shown in Figure \ref{fig:05_JT_reprod}, reproduces the fifth panel of Figure 3 in \citetalias{2010MNRAS.401..977J}.

The local minimum at $\sim1$ pc corresponds to the Jacobi radius of the binary systems ($r_J = 1.7$ pc in this setup), and roughly divides the histogram into bound systems at smaller separations and unbound systems at larger separations. The second peak\footnote{It is slightly misleading to refer to this as a peak, since it only appears as a peak because the x-axis of the Figure is logarithmic. In reality, once the binary system is unbound, the separation grows as roughly $r\sim v\Delta t \propto (\Delta t)^{1.5}$ \citepalias{2010MNRAS.401..977J}.} whose centroid lies at roughly 200 pc represents binary systems that have been unbound and are slowly drifting apart due to the small difference in their orbital frequencies.

\section{Simulations and Analysis of Results}
\label{sec:sims}

We now turn to studying how the addition of a rotating bar to the Milky Way's potential alters the final distribution of disrupted binaries compared to the axisymmetric case shown in Figure \ref{fig:05_JT_reprod}. We begin in Section \ref{subsec:sims_local} by replicating the setup used to produce Figure \ref{fig:05_JT_reprod} on a series of orbits in the vicinity of the bar's corotation resonance, and analyzing the differences between the final distributions of binary separations on the different orbits. In Section \ref{subsec:sims_full} we generalize these findings by looking at a simulation involving binary systems on a continuum of disk-like orbits, and studying how the effects of the bar's corotation resonance can be detected in local patches of the simulation. In Section \ref{subsec:sims_LM} we compare the results from two different prescriptions for the barred potential, and in Section \ref{subsec:sims_diff} we test whether orbital diffusion makes detecting the effects of the bar more challenging.

\subsection{Binary Distributions on Specific Orbits}
\label{subsec:sims_local}

Our primary simulation setup involves the quadrupole bar described in equations \ref{eq:quad_bar}-\ref{eq:bar_amp}, with a constant pattern speed: $\Omega_\mathrm{p} = 40$ km s$^{-1}$ kpc$^{-1}$. The corotation radius in this setup is $5.75$ kpc, and the half-width of the corotation resonance translates to approximately 2 kpc at its widest point for orbits with small values of $J_R$ (see the left-hand panel of Figure \ref{fig:04_pot_comp}). We begin by initializing six simulations in the barred potential, each with $5\times10^4$ binary systems on circular orbits (as defined in the unperturbed axisymmetric potential), similar to the setup described at the end of Section \ref{subsec:binary_kicks} and in Figure \ref{fig:05_JT_reprod}. In each of the six simulations, the binary systems are all launched at a random time from a specific point with respect to the bar, and allowed to evolve until the end of the simulation ($t=10$ Gyr), at which point the final positions and velocities are recorded, as well as the projected separations of each binary system (regardless of whether it became unbound during the simulation). The six points from which binary systems are launched are as follows:
\begin{enumerate}
    \item{The stable equilibrium, at $J_\phi=1275$ kpc km s$^{-1}$ and $\phi=0$.}
    \item{A representative trapped orbit, at $J_\phi=1425$ kpc km s$^{-1}$ and $\phi=0$.}
    \item{A separatrix orbit, at $J_\phi=1575$ kpc km s$^{-1}$ and $\phi=0$.}
    \item{The unstable equilibrium, at $J_\phi=1325$ kpc km s$^{-1}$ and $\phi=\pi/2$.}
    \item{A representative circulating orbit, at $J_\phi=1650$ kpc km s$^{-1}$ and $\phi=0$.}
    \item{A Solar-like orbit (i.e., a circular orbit at approximately $r=8$ kpc), at $J_\phi=1850$ kpc km s$^{-1}$ and $\phi=0$}
\end{enumerate}

The bar lies along the $y$-axis (as shown in Figure \ref{fig:04_pot_comp}), such that the stable equilibria are at $\phi=0$ and $\phi=\pi$. It is also worth repeating that in the barred potential, $J_\phi$ (or $L_z$) will not remain constant along any of these orbits, and will instead oscillate as detailed in equations \ref{eq:slow_action-angle} and \ref{eq:pendulum}; indeed, setups 3. and 4. simulate binary systems on the same orbit, but with a different initial phase. However, as initial conditions at a specific $\phi$, $J_\phi$ is a well-defined quantity, and we choose to label the orbits thus because $J_\phi$ is not a potential-dependent quantity and can be calculated directly from the observational data.

\begin{figure*}[h]
    \includegraphics[width=\textwidth]{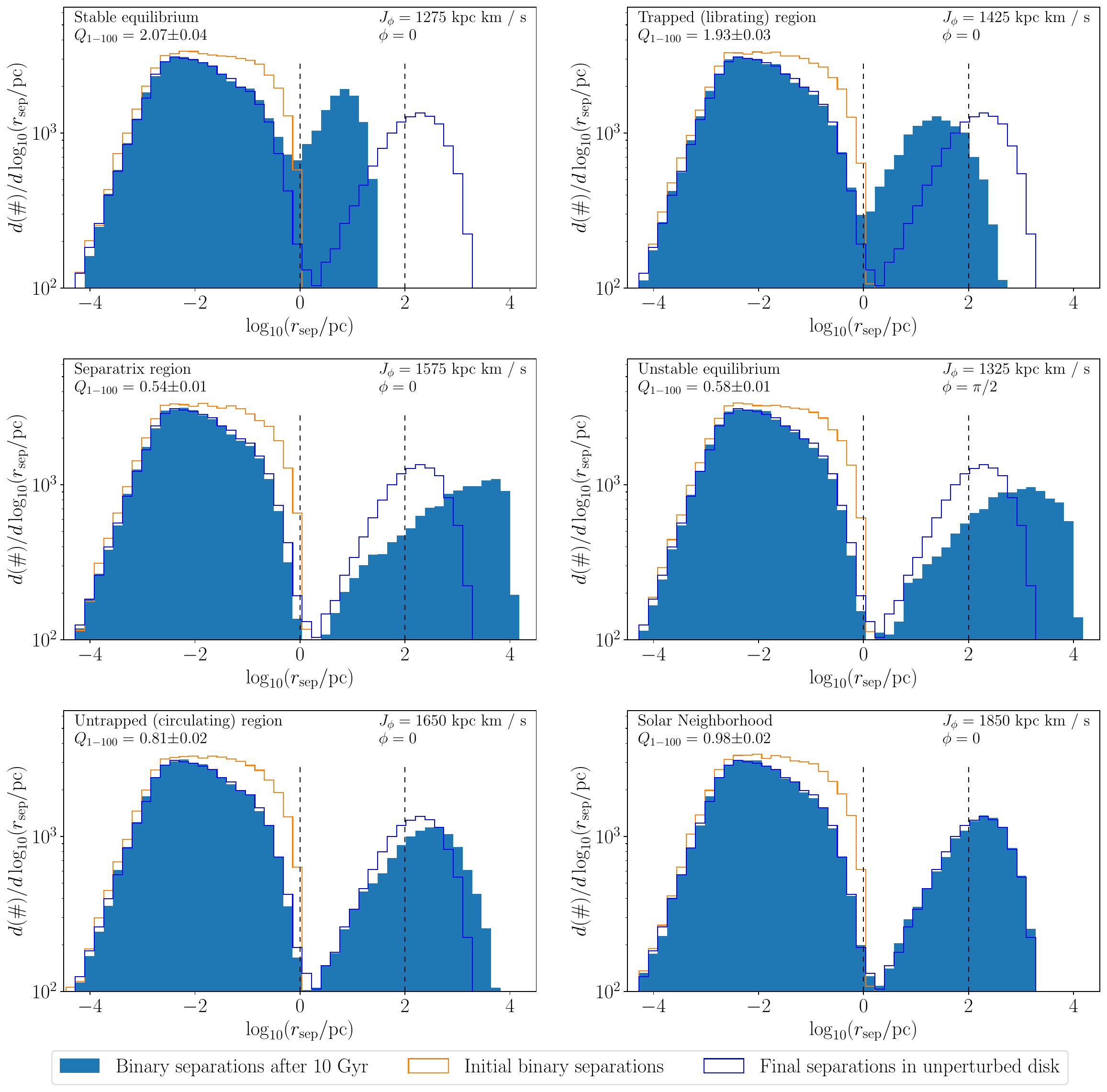}
    \caption{Histograms depicting the final projected separations of $5\times10^4$ binary systems in a barred galactic potential. Each panel corresponds to binary systems on a different Galactic orbit, as specified by the initial conditions listed at the top right of each panel (each binary is initialized on what would have been a circular orbit in the unperturbed potential). As in Figure \ref{fig:05_JT_reprod}, the initial distribution of the projected binary separations is shown in the hollow orange histogram, and the final projected separations from the unperturbed potential are shown in the hollow blue histogram for reference. The various panels exhibit clear differences in terms of the distribution of unbound binaries, with those closer to the stable balance point (top two panels) remaining at relatively small separations, while those near the separatrix (middle two panels) reach much larger separations compared to the unperturbed case. The two dashed black lines delineate the intermediate separations within which $Q_{1-100}$ is calculated.}
    \label{fig:06_localized_hists}
\end{figure*}

The final projected separations of these simulations are shown in Figure \ref{fig:06_localized_hists}, with the final projected separations of the unperturbed case from Figure \ref{fig:05_JT_reprod} shown in the blue hollow histogram for comparison.

\begin{figure*}
    \includegraphics[width=\textwidth]{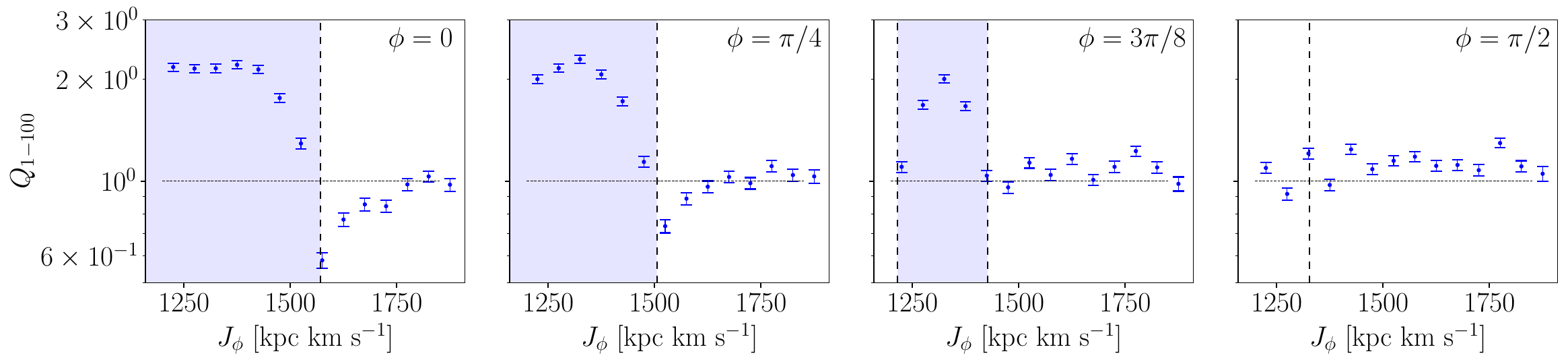}
    \caption{The excess probability of finding stars from disrupted binary systems at separations of between 1 pc and 100 pc as a function of $J_\phi$, for four different patches centered at different azimuths. A clear excess is found in the librating regions of the corotation resonance, while the separatrix is marked by a reduced probability in the first two panels. The blue shaded region represents the extent of the librating region at each azimuth (with $J_R=0$), and the separatrix is represented by the dashed black lines. Recent studies suggest the Galactic Bar is tilted by approximately 25-30 degrees with respect to the Sun's location in the Galaxy \citep[e.g.,][]{2017MNRAS.465.1621P,2019MNRAS.490.4740B}, in which case the third panel best represents the view from our vantage point.}
    \label{fig:07_P1-100_simple_bar}
\end{figure*}

The evolution of the binary systems closely resembles the expectations from the toy model described in Section \ref{subsec:toymodel}: binary systems on librating orbits close to the stable equilibrium remain close to each other even after they become unbound (top two panels), while binary systems on the separatrix diverge from each other rapidly, with many reaching separations on the scale of the Galaxy itself (middle two panels). Far from the resonance (bottom-right panel), the binary distribution is unaffected by the pendulum dynamics described earlier, and therefore it resembles the axisymmetric case. The centroid of the secondary, unbound `peak' in the projected separation distribution varies by nearly three orders of magnitude, between $<10$ pc in the librating region to $>1$ kpc near the separatrix.

In theory, these differences could allow one to indirectly identify the location and size of corotation, and thus place constraints on the pattern speed and the amplitude of the Galactic Bar. Regions in which stars from disrupted binary systems are statistically further apart from each other indicate the existence of a separatrix, whereas locations in which stars from disrupted binary systems are closer together are likely to be within the trapped region of the resonant orbit family.

In practice, however, the histograms shown in Figure \ref{fig:06_localized_hists} cannot be directly reconstructed from observational data. First and foremost -- there is no way to \textit{a priori} identify two stars that were born in a binary system which has since been disrupted, especially not in the case of two stars separated by $>100$ pc, such as those shown in the middle plots of Figure \ref{fig:06_localized_hists}.

To alleviate this challenge, we focus on stars at intermediate separations $1$ pc $ < r_\mathrm{sep} < 100$ pc, as delineated in Figure \ref{fig:06_localized_hists} by the two vertical black dashed lines. At these separations, and with the help of statistical methods and/or additional information such as metallicities or age, one might plausibly be able to identify stars that originated from the same binary system. Figure \ref{fig:06_localized_hists} shows that the cumulative number of stars at these separations is sensitive to the location of the binary systems with respect to the resonance. We define the following ratio:

\begin{equation}
    \label{eq:Q_1-100}
    Q_{1-100} \equiv \frac{P_\mathrm{perturbed} (1-100\ \mathrm{pc})}{P_\mathrm{unperturbed} (1-100\ \mathrm{pc})} \ ,
\end{equation}
where $P(1-100\ \mathrm{pc})$ is the integral of the probability distribution function of binary separation from 1 pc to 100 pc. In other words, $Q_{1-100}$ represents the excess probability of finding stars from disrupted binaries at separations between 1 pc and 100 pc in the perturbed potential compared to the unperturbed potential. The values of $Q_{1-100}$ for each of the six orbits investigated above are listed in the top right corner of the respective panels in Figure \ref{fig:06_localized_hists}: at the stable equilibrium, disrupted binaries are twice as likely to be found in the range of 1-100 pc compared to the unperturbed potential (and also compared to locations in the perturbed potential that are far from a resonance), whereas near the separatrix, disrupted binaries are only about half as likely to be found within that range of separations.

\subsection{Binary Distributions in Localized Patches}
\label{subsec:sims_full}

Besides the challenge of identifying populations of disrupted binaries at such large separation, Figure \ref{fig:06_localized_hists} makes two additional unrealistic assumptions. First, it requires knowledge of the 6D phase space information of stars across the entire Milky Way disk, while in reality we only have access to this information for a highly localized patch of the Galaxy. Furthermore, each panel in Figure \ref{fig:06_localized_hists} includes only binary systems on a specific orbit in the perturbed potential, assuming perfect knowledge of the orbital structure of the potential, even though the perturbation should of course be treated as an unknown in this problem (indeed, the perturbed potential is ultimately what we are trying to constrain using the data).

In order to study a more realistic setup, we simulate $10^6$ binary systems in the perturbed potential (using the rotating quadrupole bar as the perturbation) on a range of initial disk-like orbits. We further initialize each binary system such that the centers of mass have an initial velocity dispersion of $\sigma=10$ km s$^{-1}$. The remaining initial conditions for the binary systems themselves remain as originally described in Section \ref{subsec:binary_inits}, and as before the systems are evolved in the perturbed potential for up to 10 Gyr. An identical simulation with $10^6$ binary systems is performed in the unperturbed potential for comparison.

After logging the final separations for each binary system, we select only stars found at an azimuth of $\phi=0 \pm \pi/16$ and a Galactic radius ranging from 5.5 kpc to 8.5 kpc. We then divide this subset into bins based on the value of $J_\phi$, spaced out by $\Delta J_\phi = 50$ kpc km s$^{-1}$ (and recalling that $J_\phi = L_z$ is calculated directly from the phase space data and does not require knowledge of the Galactic potential). Finally, we calculate $Q_{1-100}$ for the ensemble of binary systems in each bin. We repeat this procedure for patches centered on three other azimuthal positions: $\pi/4$, $3\pi/8$, and $\pi/2$. The results are shown in Figure \ref{fig:07_P1-100_simple_bar}, with each panel corresponding to one of the azimuthal patches. Many recent studies agree that the Milky Way's Galactic Bar is tilted by $25-30$ degrees with respect to the Sun \citep[e.g.,][]{2017MNRAS.465.1621P,2019MNRAS.490.4740B}. If true, the view from our vantage point would most resemble the third panel in Figure \ref{fig:07_P1-100_simple_bar}.

\begin{figure*}
    \includegraphics[width=\textwidth]{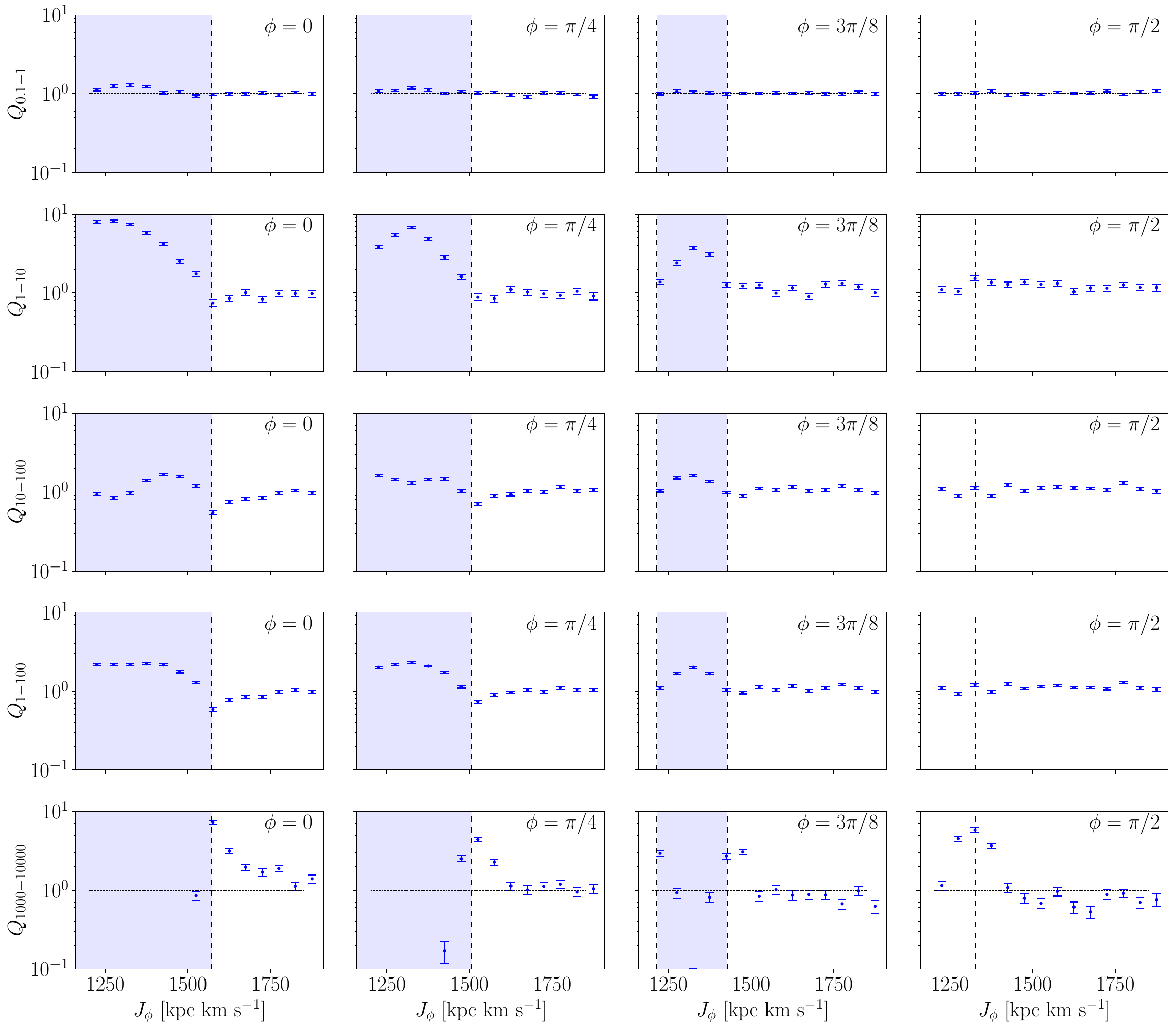}
    \caption{Same as Figure \ref{fig:07_P1-100_simple_bar} for different values of the minimum and maximum separations used to calculate $Q$ (Equation \ref{eq:Q_1-100}). As in Figure \ref{fig:07_P1-100_simple_bar}, each column corresponds to a different azimuthal patch, and the rows correspond to the excess probability of finding stars from binary systems between (1) 0.1 pc and 1 pc, (2) 1 pc and 10 pc, (3) 10 pc and 100 pc, (4) 1 pc and 100 pc, and (5) 1 kpc and 10 kpc. Note that the $y$-axis spans a much larger range in this figure than it does in Figure \ref{fig:07_P1-100_simple_bar}. The data in the fourth row is identical to Figure \ref{fig:07_P1-100_simple_bar}, and is shown again to facilitate the comparison.}
    \label{fig:08_Prange}
\end{figure*}

The location of the corotation resonance is clearly visible in the first three panels, with nearly double the likelihood of finding the stars from disrupted binaries at separations of up to 100 pc within the librating regime. Due to the geometry of the corotation resonance, the width of the librating regime is at a maximum at $\phi=0$ and shrinks to zero at $\phi=\pi/2$. In the first two panels, the separatrix region is also clearly distinguishable as an area where disrupted binaries separated by 1-100 pc are \textit{less} likely. At exactly $\phi=\pi/2$ the signal from the resonance essentially disappears, not unlike the relatively uniform color across all values of $I$ at $\varphi = \pm\pi$ in Figure \ref{fig:03_Phase_Space_of_a_Pendulum_from_binary_separations}.

\begin{figure*}
    \includegraphics[width=\textwidth]{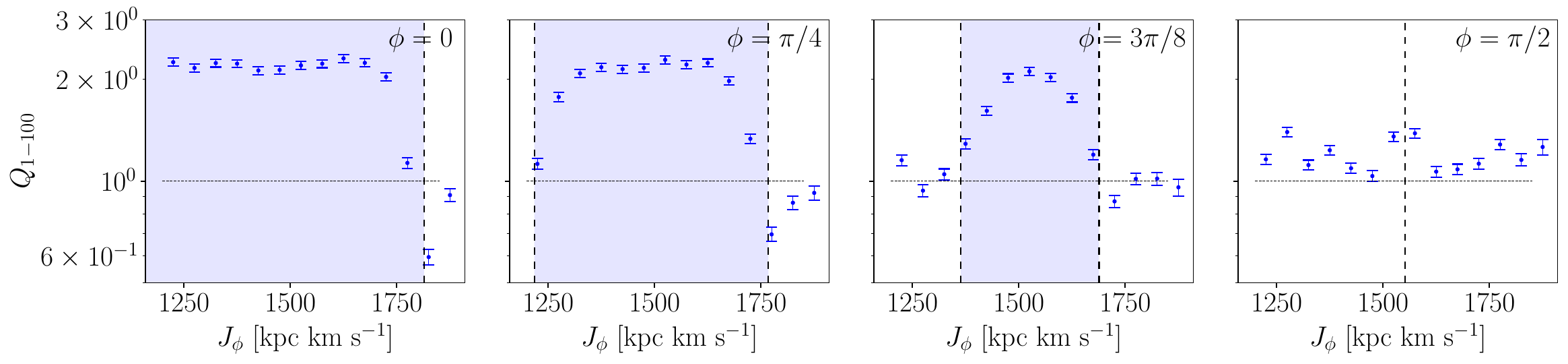}
    \caption{The equivalent of Figure \ref{fig:07_P1-100_simple_bar} in the \citet{1992ApJ...397...44L} bar potential. The center and width of the corotation resonance is shifted slightly outward, compared to the quadrupole bar potential used to produce the previous figure. Notably, the width of the region in which stars are less likely to be found at separations of 1-100 pc (corresponding to the separatrix), is not any larger than in Figure \ref{fig:07_P1-100_simple_bar}, even though the \citet{1992ApJ...397...44L} potential supports a sizable band of chaotic orbits with Lyapunov times shorter than 10 Gyr at the separatrix. We conclude that chaos is not a necessary ingredient for understanding the effect studied here.}
    \label{fig:09_P1-100_LM}
\end{figure*}

Figure \ref{fig:07_P1-100_simple_bar} includes a dashed line at $Q_{1-100} = 1$ to highlight the excess probability in the libration region and the lack thereof in the separatrix region. However, we note that the figure is equally helpful for determining the location and width of the resonance even in the absence of this reference line, because the key information lies in the variation of the signal as a function of the orbital angular momentum. One could arrive at the same result by defining $Q_{1-100}$ with respect to the binary separation probability distribution function at the Solar neighborhood (which in this example happens to be located far enough away from the resonance such that $Q_{1-100}=1$ anyway, though this need not be the case).

A full face-on view of the Galactic Disk is shown in the leftmost panel of Figure \ref{fig:12_face_on_disk}, colored by the local value of $Q_{1-100}$ at each pixel (the other two panels in the figure represent simulations discussed in the sections below). As in Figure \ref{fig:07_P1-100_simple_bar}, the resonant regions appear clearly, and can be compared directly to the shape and size of the near-separatrix orbit shown in the left-hand panel Figure \ref{fig:04_pot_comp}.

The choice of $Q_{1-100}$ is of course arbitrary, and in many situations not the best option. Figure \ref{fig:08_Prange} demonstrates how various choices of minimum and maximum separations for Equation \ref{eq:Q_1-100} contain different information. At very small separations (i.e., less than the Jacobi radius), $Q_{0.1-1}$ shows very little variance across the disk, as expected. $Q_{1-10}$ is highly sensitive to the central region of the resonance, but does not show as significant a drop at the separatrix. $Q_{10-100}$ does exhibit the expected drop at the separatrix, but also dips down at the center of the librating region, exactly as one might expect based on the top-left panel in Figure \ref{fig:06_localized_hists}. Finally, $Q_{1000-10000}$ is perhaps the most telling, with a clear deficit within the librating region, and a large excess at the separatrix. However, as mentioned above, identifying that two stars at such large separations were originally part of the same binary system is highly challenging.

\subsection{Long \& Murali Bar}
\label{subsec:sims_LM}

One drawback of using the quadrupole prescription is that it produces a very `orderly' series of resonances. A more realistic bar is likely to have higher order moments with different amplitudes that source their own spectrum of resonances. Though the resonances from the $m=2$ component of the bar still dominate, the boundaries of the resonances become blurred by a band of chaotic orbits that forms around the separatrices.

To investigate how this affects the distribution of disrupted binaries, we repeat the $10^6$ binary system experiment described in Section \ref{subsec:sims_full}, this time with a \citet{1992ApJ...397...44L} bar instead of the quadrupole bar. For the sake of consistency, we retain the bar's pattern speed from the previous section ($\Omega_\mathrm{p} = 40$ km s$^{-1}$ kpc$^{-1}$), noting that the location of the corotation resonance moves slightly outward in this prescription due to the increased total mass of the halo.

Figure \ref{fig:09_P1-100_LM} is the equivalent of Figure \ref{fig:07_P1-100_simple_bar} for this potential. As noted above, the location of the stable equilibrium point is now further out (at $\sim1500$ kpc km s$^{-1}$ instead of $\sim1300$ kpc km s$^{-1}$). The librating region is also clearly wider than in the previous configuration -- this is purely due to the choice of parameters, and using a lower value for the mass of the bar would shrink the resonant region. The rightmost panel in Figure \ref{fig:12_face_on_disk} also depicts these characteristics for the face-on view of the disk with the \citet{1992ApJ...397...44L} bar.

A key difference between the \citet{1992ApJ...397...44L} potential and the quadrupole bar used in the previous sections is that the former generates a collection of chaotic orbits in a band surrounding the separatrix, with Lyapunov times shorter than 10 Gyr. However, this has at most a secondary effect on the eventual distribution of the disrupted binaries in the system: the region of lower probability around the separatrix is not any wider or more pronounced than the one in Figure \ref{fig:07_P1-100_simple_bar}.

A likely explanation for this is that the frequency difference between two stars from a disrupted binary system as one nears the separatrix is large enough to cause those stars to reach separations on the scale of the entire orbit. As such, even though this frequency difference is a smooth and predictable function, its effects (in terms of the overall separation between two stars from a disrupted binary system) are not greatly enhanced by the existence of chaotic orbits. We therefore conclude that chaos is not a necessary ingredient of the phenomenon described here: the distribution of binary separations varies greatly even in the absence of chaotic orbits, and the existence of chaotic orbits with relatively short Lyapunov times does not greatly enhance the observable variation.

It is reasonable to imagine that a more significant region of chaotic orbits, such as that sourced by resonance overlap between two non-axisymmetric features with different pattern speeds (i.e., the Galactic Bar and spiral arms), would have a more pronounced effect, created a wider region or even a plateau with $Q_{1-100} < 1$. We leave this as a topic for future investigation.

\begin{figure}
    \includegraphics[width=0.95\columnwidth]{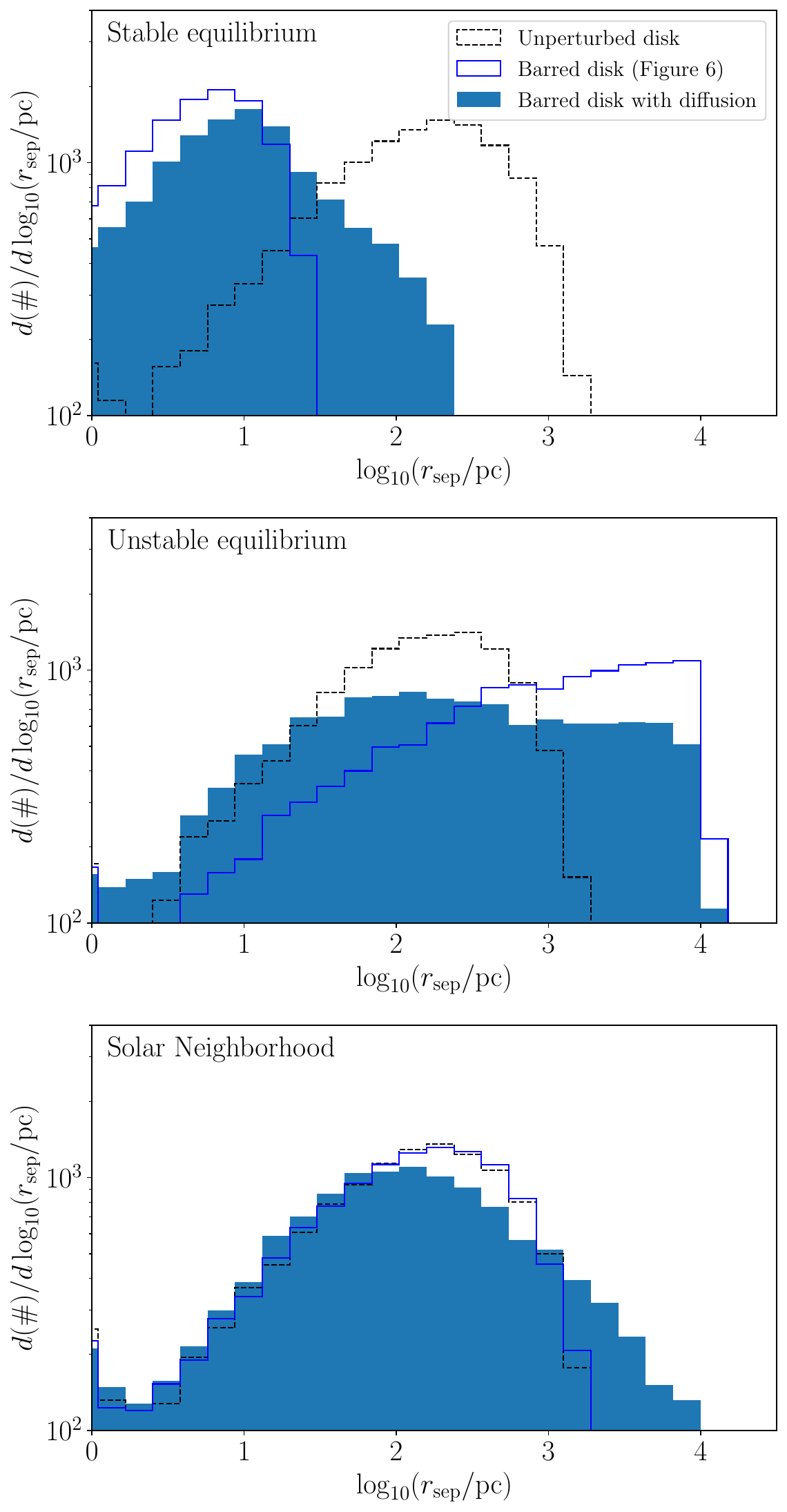}
    \caption{Histograms depicting the final projected separations of binary systems in a barred galactic potential with orbital diffusion. For each initial orbit, the barred disk with diffusion (filled histogram) is compared to the barred disk without diffusion (solid blue line) shown in Figure \ref{fig:06_localized_hists}. The unperturbed disk histograms from Figure \ref{fig:06_localized_hists} are also shown for reference (black dashed lines). The three panels here correspond to the top-left, middle-right, and bottom-right panels of Figure \ref{fig:06_localized_hists}, respectively. Orbital diffusion tends to moderate the effect of the bar resonance, moving the average projected separation to higher values at the librating zone and to lower values near the separatrix. The plot only extends to a minimum projected separation of 1 pc, since the distribution of bound binaries is largely unaffected.}
    \label{fig:10_localized_hists_diff}
\end{figure}

\subsection{Accounting for Orbital Diffusion}
\label{subsec:sims_diff}

The orbits of two stars that originated in the same binary system evolve over time as a result of kicks imparted by a variety of perturbations that are ever-present in the Galaxy. To first order, the nature of these kicks is determined by the scale of the perturbation; large-scale perturbations, such as passing satellite galaxies or the Galactic spiral structure, impart coherent kicks to both stars that can greatly alter the center of mass motion of the binary system while barely inducing any tidal forcing. Small-scale perturbations, such as individual stars that pass close by the binary system, do the opposite, by imparting small but different kicks to each star, that contribute mostly to the tidal forcing.

Thus far, the focus here has been on the small-scale perturbations (as in \citetalias{2010MNRAS.401..977J}), because these perturbations are the primary drivers of a binary system's disruption. However, the larger perturbations are also important in the context of this work, because they can drive the center of mass of the binary system to different locations with respect to the bar resonance (even after the binary system has been tidally disrupted).

In order to investigate how these larger scale perturbations affect the distribution of disrupted binaries at large separations, we repeat all the simulations described in Section \ref{subsec:sims_local}-\ref{subsec:sims_full} with an added step for incorporating coherent kicks. As described in Section \ref{subsec:binary_kicks}, we impart a kick to each of the binary's stars every $\Delta t_\mathrm{p}$. We now include an additional and coherent kick to both stars (i.e., the same velocity is added to both stars), such that by the end of the simulation, the velocity dispersion of the full population has grown by $\sim20$ km/s.

\begin{figure*}
    \includegraphics[width=\textwidth]{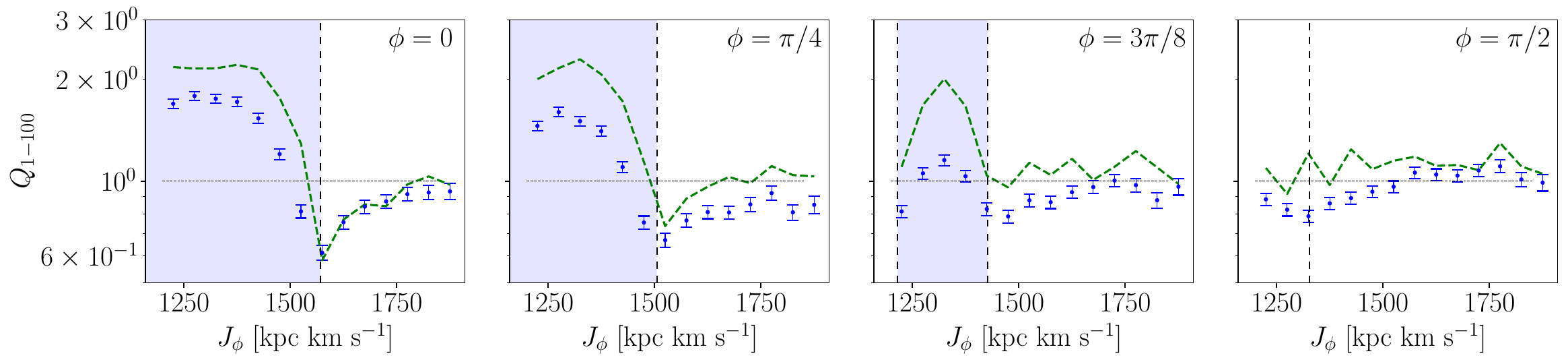}
    \caption{The equivalent of Figure \ref{fig:07_P1-100_simple_bar}, accounting for orbital diffusion through the addition of periodic coherent kicks to the binary systems in the simulation (the kicks are included in both the perturbed and unperturbed simulations used to calculate the value of $Q_{1-100}$, such that the only difference between the two simulations remains the presence of the bar). The setup used to produce this figure includes a quadrupole bar identical to the one used in the simulation for Figure \ref{fig:07_P1-100_simple_bar}, and the black dashed lines in each panel represent the value of $Q_{1-100}$ from that figure. Besides reducing the probability globally, diffusion tends to moderate the effects of the resonance, leading to less sharp and somewhat wider features.}
    \label{fig:11_P1-100_diff}
\end{figure*}

Figure \ref{fig:10_localized_hists_diff} compares the eventual distribution of $5\times10^4$ binary systems after 10 Gyr for three different initial orbits (the hollow blue histograms in the three panels are identical to the top left, middle right, and bottom right panels of Figure \ref{fig:06_localized_hists}). Diffusion appears to moderate the effect of the resonant region: pushing disrupted stars from binaries initialized at the stable equilibrium to larger separations, while moving the distribution at the unstable equilibrium to smaller separations than the simulation with no coherent diffusion. At locations far from the resonance, the distribution is biased slightly to higher separations, leading to a global reduction in the value of $Q_{1-100}$.

This suggests that diffusion has a smoothing effect on the structure variation shown in Figures \ref{fig:07_P1-100_simple_bar} and \ref{fig:09_P1-100_LM} -- leading to wider, if less sharp, features. This is shown in Figure \ref{fig:11_P1-100_diff} -- the peaks at the stable equilibrium and the troughs at the separatrix are seen at a reduced contrast, while the variations in $Q_{1-100}$ are broader and smoother. In some instances, the separatrix actually becomes marginally more noticeable due to this broadening, in particular in the third panel centered on $\phi=3\pi/8$ (which may be particularly relevant, given it is the panel that likely depicts the view from our vantage point in the Galaxy).

\begin{figure*}
    \includegraphics[width=\textwidth]{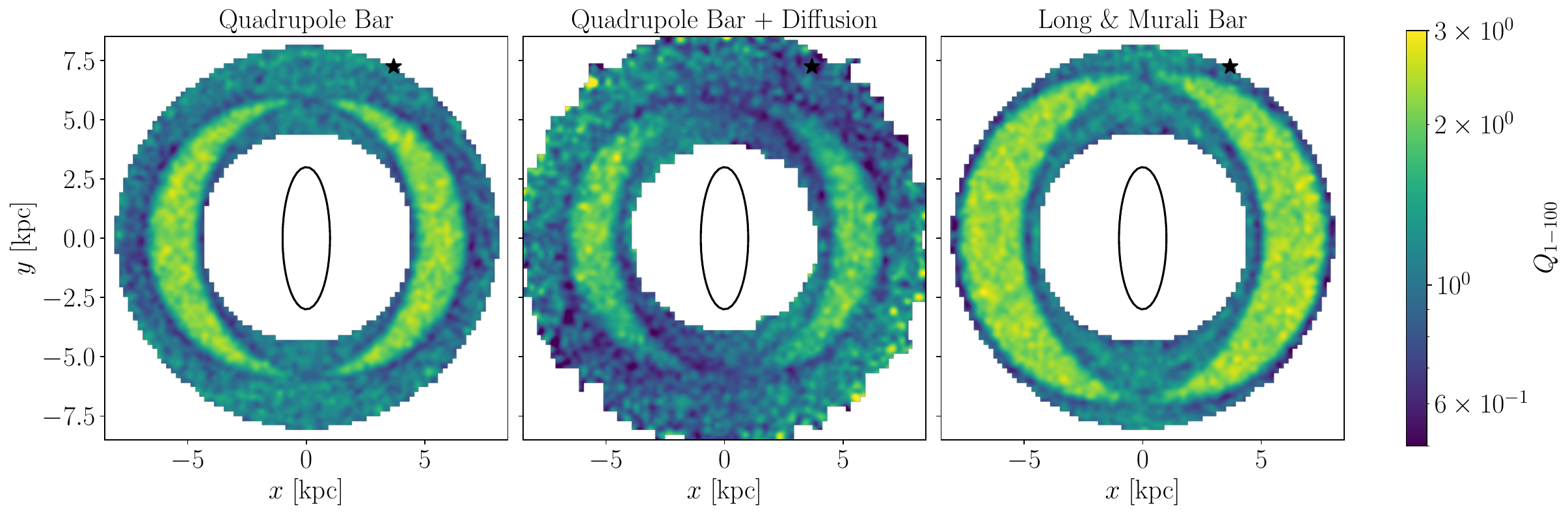}
    \caption{A face-on view of the Galactic Disk in each of the three simulations described in Sections \ref{subsec:sims_full}-\ref{subsec:sims_diff}, colored by the localized value of $Q_{1-100}$. The orientation of the bar is depicted using the black ellipse (the size and shape of the ellipse do not correspond to any of its physical characteristics besides its pitch angle), and the black star represents the consensus view on the approximate location of the Sun with respect to the bar. The regions affected by the corotation resonance are clearly visible in each of the three panels, appearing as islands with high values of $Q_{1-100}$ at their centers, surrounded by a lower-value band at their outskirts. As mentioned in Section \ref{subsec:sims_diff}, diffusion has a smoothing effect on the features shown in the middle panel, while slightly widening the bands of low $Q_{1-100}$ near the separatrix. The middle panel also depicts a slight asymmetry, similar to the asymmetric structures discussed in \citet{2023ApJ...954...12H}.}
    \label{fig:12_face_on_disk}
\end{figure*}

The effects of adding orbital diffusion are similarly apparent in the comparison of the leftmost panel of Figure \ref{fig:12_face_on_disk} to the middle panel. The boundaries of the resonant regions are identical across the two panels (because the pattern speed and amplitude of the bar are the same). However, the middle panel, which corresponds to the simulation with the added orbital diffusion described in this section, includes features that are somewhat `smeared out' and not as well-defined as those of the lefttmost panel.

\citet{2023ApJ...954...12H} showed that orbital dynamics near galactic resonances (and specifically at the corotation resonance with a Galactic Bar) can be significantly affected as a result of orbital diffusion. The attenuation of the variation in binary distributions shown in Figure \ref{fig:11_P1-100_diff} is a manifestation of this prediction, though in this case the diffusion is clearly not powerful enough to erase the signal entirely. In the notation of \citet{2023ApJ...954...12H}, the setup of our simulation is equivalent to $\Delta\sim0.1$, where $\Delta$ is the diffusion strength, which is the ratio of the resonance's libration time to the time it takes to diffuse across the resonance. A higher value of $\Delta$ would lead to additional attenuation of the signal, and, eventually, to its complete erasure. However, typical estimates of $\Delta$ in the Milky Way are not expected to be significantly greater than $0.1$, and thus are not expected to completely destroy the signal discussed here.

\section{Discussion and Conclusions}
\label{sec:disc}

This work has focused on the tidal disruption and subsequent evolution of binary systems embedded in galactic potentials, with a specific emphasis on the effects of non-axisymmetric potential perturbations such as those sourced by galactic bars. Much of this work follows the approach outlined in \citetalias{2010MNRAS.401..977J}, who were the first to follow the evolution of stars from binary systems after the disruption of those systems by the Galactic tidal field, and predict the eventual distribution of projected separations. The primary goal here has been to extend the treatment in \citetalias{2010MNRAS.401..977J} in order to study the variation in the distribution of disrupted binary systems due to the presence of a rotating galactic bar. In particular, we have demonstrated that:
\begin{enumerate}[(i)]
    \item As predicted in \citetalias{2010MNRAS.401..977J}, the distribution of projected separations forms a secondary peak (when plotted in \# per unit log separation). The centroid of this peak is strongly affected by the characteristics of the Galactic potential, and varies by nearly three order of magnitude, as a function of the location of the binary systems with respect to resonant orbit families in the Galactic potential.
    \item In librating regions, stars from disrupted binary systems stay close to one another long after the system is tidally disrupted, because the libration frequencies near the stable equilibrium of the resonance are nearly identical across a large range of separation in phase space.
    \item Near the edge of the resonance, the stars drift apart from each other much faster due to the large frequency gradient close to the separatrix. In some cases, they are able to reach separations comparable to the size of the Galaxy itself within the Galactic lifetime.
    \item Measuring the likelihood of finding stars at intermediate separations (1-100 pc) reveals order unity variations in the local structure that can be used to identify the location and width of the resonance. For a rotating bar, these can be translated directly into the pattern speed and the amplitude of the bar, assuming knowledge of the background (axisymmetric) potential.
    \item The presence of orbital diffusion can attenuate and `smear' the variation in the likelihood of finding disrupted binaries at specified separations. For the expected values of the diffusion coefficients in the Galaxy, this is likely to somewhat reduce the signal described above, though not to erase it entirely.
\end{enumerate}

Galaxies are complex systems with many time-dependent perturbations that complicate and alter the picture described above. These include the presence of additional structures such as spiral arms, GMCs, substructure from various theorized models of dark matter, and other external perturbers such as infalling dwarf galaxies. Moreover, throughout this work we have assumed constant parameters for the Galactic Bar, where in reality the bar is most likely both growing in amplitude and slowing down. Dealing with these complications quantitatively is beyond the scope of the present work, but we can use the intuition built from the simplified model presented above to qualitatively discuss the effects of a few of these.

The presence of an additional time-dependent perturbation with a different frequency -- say, the Galaxy's spiral arms -- will create other resonant orbit families. In certain situations, the resonances from different perturbations will overlap, leading to large swaths of chaotic orbits. Even if the resonances do not overlap, this can still create large bands of chaotic orbits around the separatrices of each orbit family (considerably larger than the chaotic band briefly mentioned in Section \ref{subsec:sims_LM}). In these situations, a plot like Figure \ref{fig:07_P1-100_simple_bar} would likely reveal fewer areas of higher likelihood, and broader areas of lower likelihood. In particular, it is reasonable to assume that large plateaus of low likelihood would be present around the separatrices, indicating large bands of chaotic orbits.

The evolution of the pattern speed and amplitude of the bar (or of spiral arms, or any other large and periodic Galactic structures) can also alter the predictions described above. If, for example, the bar's rotation is rapidly slowing down (on a timescale comparable to the libration time: $2\pi / \omega_{\ell,\mathrm{res}}$), the resonant effects that drive the variations discussed above do not have time to manifest \citep[see also][]{2021MNRAS.500.4710C}, leading to attenuation or even complete erasure of the variations described above. However, based on other estimates of the bar's evolution, the rate of growth and slowdown is expected to be sufficiently low to allow the resonant effects to take hold \citep[see, e.g.,][]{2021MNRAS.505.2412C}. We expect the distribution of projected separations to be mostly sensitive to the current characteristics of the bar. The attenuation due to diffusion described in Section \ref{subsec:sims_diff} is also likely more important than the effects of a gradually slowing or growing bar.

In addition to the aforementioned complications, our model makes a series of simplifying assumptions regarding the observed signal due to disrupted binaries that may be worth revisiting. First, it is important to note that the quantitative values for the binary separation probability distribution function should not be taken as predictive, because they are heavily dependent on the initial conditions of the binary systems. As described in Section \ref{subsec:binary_inits}, our setup involves a series of simplifying assumptions regarding the binary initial conditions, many of which are likely inaccurate \citep[see, e.g.,][]{2023ApJ...955..134R}. Nevertheless, the main argument we have made throughout this work is that while the absolute value of the distribution function may be an unknown for a variety of reasons, including uncertainties regarding the binary initial conditions, one can still use the \textit{variation} of the observed distribution to detect and characterize Galactic structure.

This, however, assumes that the initial conditions of binary systems, whatever they may be, are independent of the location in the Galaxy, and specifically of the location with respect to resonances. It is possible, and perhaps even likely, that this is not the case, and that the star formation rate is dependent on (or at least correlated with) the location of bar resonances. For instance, NGC 1398 provides a compelling visual case for a galaxy in which the star formation rate may be considerably higher at the outskirts of the corotation resonance than in its interior. With that in mind, and discounting the unlikely scenario in which these two effects exactly cancel each other out, studying the variations in the eventual distribution of disrupted binary systems may still shed interesting light on the Galaxy's resonant structure, even if the variations take on a different form compared to those presented in Section \ref{sec:sims}.

One of the most promising approaches for detecting stars from disrupted binaries at separations between 1-100 pc lies in measuring the two-point correlation function (2PCF) in the Gaia dataset. Previous measurements of the local 2PCF, including \citet{2019MNRAS.484.2556L}, \citet{2021ApJ...922...49K}, and \citet{2023ApJ...942...41H}, have studied the global signal (e.g., across the entire Gaia footprint), rather than searching for variations that are indicative of Galactic structure, as predicted in this work. Having laid out the theoretical basis here, we leave a detailed investigation of the Gaia 2PCF for future work.

Finally, it is worth noting that in their study of the Gaia 2PCF, \citet{2021ApJ...922...49K} reported that they were unable to detect a signal that could be directly attributed to disrupted wide binaries, as predicted in \citetalias{2010MNRAS.401..977J}. A likely reason for this is that the wide binary signal is completely overshadowed by the signal stemming from dissolving open clusters in the Galactic Disk. Importantly, though this work has focused on modeling wide binaries, the effects described here will have a very similar manifestation in the distribution of stars from disrupted clusters. The pendulum dynamics responsible for the variations reported in this work are identical to the mechanism used to explain features in stellar streams from disrupted globular clusters on near-resonant orbits \citep{2023ApJ...954..215Y}. Bar resonances have indeed been invoked to explain unique features seen in several observed streams \citep{2016ApJ...824..104P,2020ApJ...889...70B,2024arXiv240514933D}. Therefore, if dissolving open clusters are indeed the main contributors to the 2PCF reported in \citet{2021ApJ...922...49K}, we have every reason to believe that the 2PCF will exhibit spatial variations dependent on the location of bar resonances.

% \begin{acknowledgments}
\section*{Acknowledgements}

It is a pleasure to thank Sihao Cheng and Chris Hamilton for many helpful conversations that greatly contributed to this work. I also thank Rimpei Chiba, Hsiang-Chih Hwang, Kathryn Johnston, Melissa Ness, Mor Rozner, Ralph Sch\"{o}nrich, and Scott Tremaine for their comments and suggestions. TDY is supported through the Bezos Member Fund and the Fund for Natural Sciences at the Institute for Advanced Study. This research was supported in part by grant NSF PHY-2309135 to the Kavli Institute for Theoretical Physics (KITP).

% \end{acknowledgments}

\software{Astropy \citep{2013A&A...558A..33A,2018AJ....156..123A,2022ApJ...935..167A},
          Gala \citep{2017JOSS....2..388P}, 
          Matplotlib \citep{Hunter:2007}, 
          Numpy \citep{harris2020array},
          Scipy \citep{2020SciPy-NMeth},
          }

\bibliography{references,references2}{}
\bibliographystyle{aasjournal}

\end{document}